\documentclass[doublecol, letterpaper]{epl2} 
\usepackage{graphicx}
\usepackage{amsmath}
\usepackage{amssymb}
\usepackage{wasysym}

\title{Localized states in sheared electroconvection}

\author{Peichun Tsai\inst{1}, Stephen W.~Morris\inst{1}, and Zahir A. Daya\inst{2}}

\institute{
	\inst{1} Department of Physics, University of Toronto - 60 St. George St., Toronto, Ontario, Canada M5S 1A7\\
	\inst{2} Defence Research \& Development Canada - 9 Grove Street, Dartmouth, Nova Scotia, Canada B2Y 3Z7
}

\date{\today}

\abstract{ Electroconvection in a thin, sheared fluid film displays a rich sequence of bifurcations between different flow states as the driving voltage is increased.  We present a numerical study of an annular film in which a radial potential difference acts on induced surface charges to drive convection.  The film is also sheared by independently rotating the inner edge of the annulus.  This simulation models laboratory experiments on electroconvection in sheared smectic liquid crystal films. The applied shear competes with the electrical forces, resulting in oscillatory and strongly subcritical bifurcations between localized  vortex states close to onset.  At higher forcing, the flow becomes chaotic {\it via} a Ruelle-Takens-Newhouse scenario.  The simulation allows flow visualization not available in the physical experiments, and sheds light on previously observed transitions  in the current-voltage characteristics of electroconvecting smectic films.}

\begin{document}

\maketitle

Driven, dissipative nonlinear systems sometimes exhibit spatially localized structures which are analogous to solitons~\cite{localized_theory_review}.
% localized_chaos_issue}.  
Examples are found in systems as diverse as vegetation patterns~\cite{bio}, 
%wide lasers~\cite{lasers},   to save space
vibrated granular media~\cite{oscillons} and ferrofluids~\cite{localized_ferro_expt}. Such states can also arise in %very well-studied 
fluid-mechanical settings such as binary fluid convection~\cite{binary_expts, snakes_bin_fluid_conv} (where they have been called ``convectons''~\cite{convectons}), in electroconvecting nematic liquid crystals~\cite{worms}, and in very general model equations~\cite{snakes_SHE,
%snakes_large_scale_mode, 
Dawes_JFM}.  
In this Letter, we describe a new and unexpected type of localized patterns: solitary vortex states in two-dimensional, sheared electroconvection. 
% new sentence emphasizing the smallness of the states
 We show numerically that these states can be extremely localized, consisting of only a single, isolated vortex surrounded by a uniform background state.  The presence of localized states in this system is surprizing and interesting because it results from the interaction of a circular Couette shear with two-dimensional convection in an especially simple, highly symmetric geometry.

 Theses states can be experimentally realized using thin free-standing films of smectic liquid crystals~\cite{flow_vis,
 %zahir_prl, 
 DDM_PhyFluids_1999, DDM_PRE_2001, ZDaya_PRE2002, DDM_coD2_PRE05,  TDDM_simulation07}.  Here,  direct numerical simulation allows us to study the spatial structure of the full velocity, charge and potential fields~\cite{TDDM_simulation07}.  This approach compliments existing theory~\cite{DDM_PhyFluids_1999, TDDM_simulation07} and experiments~\cite{flow_vis,
 % zahir_prl, 
 DDM_PhyFluids_1999, DDM_PRE_2001, ZDaya_PRE2002, DDM_coD2_PRE05} on this system, which mainly consisted in observations of the total current through the thin film, without flow visualization. The simulation reveals localized states in the form of vortices which travel in the direction of the applied shear.  These are preceded by lower-amplitude, extended traveling and oscillatory vortex states. At sufficiently high levels of electrical forcing, the flow becomes chaotic {\it via} a Ruelle-Takens-Newhouse scenario~\cite{RTN_route_to_chaos}.  Our results serve to strongly motivate new experiments and theory to elucidate the dynamics of sheared 2D convection.  More generally, this system with naturally periodic boundary conditions, and for which forcing and shear are independently controllable, will be an interesting place to examine recent ideas on higher dimensional invariant manifolds on the way to turbulence~\cite{cvit, 
%RBraun_PRE98, FFeudel_PRE95, 
Heijst_PRL05}.
%, quasi_2D_exp_JFM1986}.
 
 %
 % remove a reference
 %
 % RBC_route_to_chaos1}.  

\begin{figure}
\begin{center}
\includegraphics[width=1.3in]{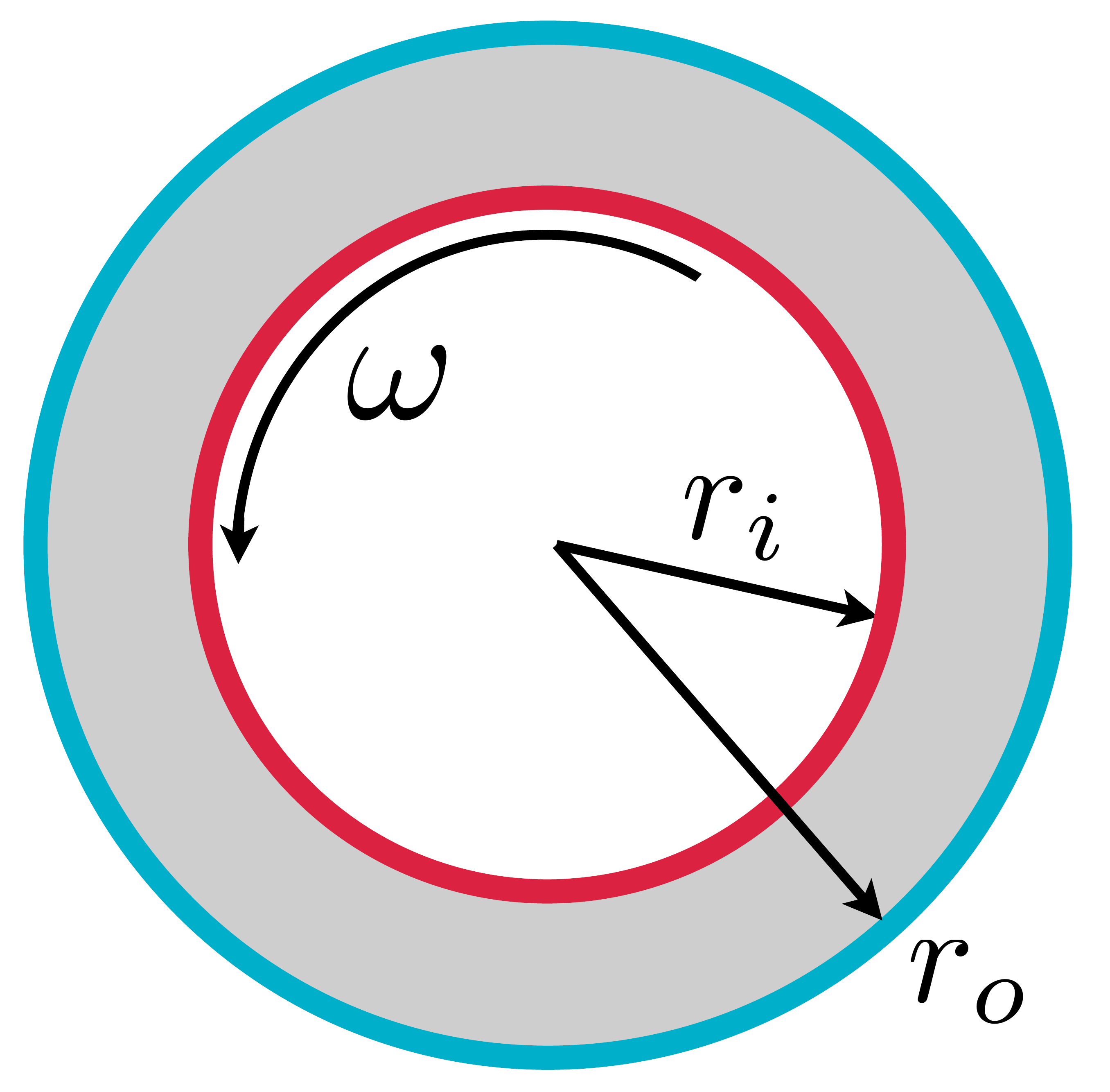}
\end{center}
\caption{\label{schematic_annulus} (Color online) Schematic of the sheared 2D annular film.}
\end{figure}

% include shear and alpha parameters here, not later
Our numerical study simulates a laboratory experiment shown schematically in Fig.~\ref{schematic_annulus}.  The system consists of a submicron-thick liquid crystal film freely suspended between concentric circular electrodes. The weakly conducting film is driven to convect when a sufficiently large electric potential is imposed across it. The inner edge of the annular film is held at absolute potential $V$ with respect to infinity, while the outer edge is held at zero potential.  In addition to the control parameter associated with this electrical driving force, it is possible to independently rotate the inner electrode, which imposes an azimuthal Couette shear on the film. The experimental signature of convection consists of measurements of the total current through the film, which is increased by convective flow.  

The film develops a surface charge configuration which is unstable to the applied voltage.  This instability is closely analogous to that of Rayleigh-B{\'e}nard convection, in which an inverted density distribution is unstable to buoyancy forces.  As in Rayleigh-B{\'e}nard convection, there are two important dimensionless parameters.  The Rayleigh-like number ${\cal R}$ describes the ratio of the electrical forcing to the viscous and electrical dissipation; this serves as the main control parameter.  The Prandtl-like number ${\cal P}$ is the ratio of charge relaxation time $\tau_q$ to viscous relaxation time  $\tau_v$.  The annular geometry of the film is characterized by the radius ratio $\alpha$. These parameters are given by~\cite{DDM_PhyFluids_1999}
\begin{equation}
{\cal R} \equiv \frac{\epsilon_0^2V^2}{\sigma \eta}, \quad  {\cal P} \equiv \frac{\epsilon_0\eta}{\rho\sigma d}  \quad \text{and} \quad \alpha \equiv \frac{r_i}{r_o},
\end{equation}
where $\rho$, $\eta$ and $\sigma$ are the 2D mass density, shear viscosity and electrical conductivity, respectively.  The width of the film is $d = r_o - r_i$, and $\epsilon_0$ is the permittivity of free space.

% ------ \section*{Sheared Convection} ---------

Rotation of the inner electrode introduces a circular Couette shear to the base state below the onset of convection.   The strength of the shear is characterized by a dimensionless shear Reynolds number \begin{equation}
{\cal R}e={\rho d \omega r_i }/{\eta} ,
\end{equation}
where $\omega$ is the angular rotation frequency of the inner electrode.  The basic equations, given in detail in Refs.~\cite{DDM_PhyFluids_1999} and \cite{TDDM_simulation07},  are nondimensionalized using the film width $d$ as the unit of length and the charge relaxation time $\tau_q = \epsilon_0 d/\sigma$ as the unit of time. Recall that  $\sigma$ is the 2D conductivity. In these units, the dimensionless rotation rate of the inner electrode is $\Omega = \omega \tau_q$.  

The dimensionless electrical current through the film is described by the Nusselt number, $Nu=I/I_0$, where $I_0$ is the fraction of the current contributed by pure conduction at the same voltage.  In a convecting film,  $Nu>1$. The quiescent film undergoes a bifurcation from conduction to convection when a voltage larger than a critical value $V_c$ is applied, corresponding to a critical value ${\cal R}_c$, which in general depends on $\alpha$,  ${\cal P}$ and ${\cal R}e$.  Typically,  ${\cal R}_c \sim 100$.  Just above onset, the flow is characterized by a number of counter-rotating vortices arranged around the annulus, filling its entire circumference.  Under shear, the vortex pattern rotates, traveling in the same sense as the inner electrode. 

% When a voltage $\apprge 100~{\cal R}_c$ is applied, the convective flow becomes turbulent.  Experiments in the turbulent regime have shown that $Nu$ exhibits a local power law scaling with ${\cal R}$ for ${\cal R} \apprge 10^4$~\cite{TDM_PRL04}. 

We generalize our previous numerical scheme~\cite{TDDM_simulation07} to encompass the case of nonzero shear.  Denoting the quantities in the base state below the onset of convection by a superscript zero, the only new nonzero quantity that appears in the perturbed equations (Eqns. 18 and 19 in Ref.~\cite{TDDM_simulation07}) 
is the radial derivative of the base state stream function $\phi^{(0)}$ given by
\begin{equation}
\frac{\partial \phi^{(0)}}{\partial r} = \frac{\alpha^2\Omega}{1-\alpha^2}\left( r-\frac{1}{r(1-\alpha)^2}\right), 
\end{equation}
where $r$ is the dimensionless radial coordinate in units of $d$.  With this change, the numerical treatment described in Ref.~\cite{TDDM_simulation07} can be carried out for the sheared case.

The direct numerical simulations used a pseudo-spectral scheme, based on the Fourier Galerkin method in the azimuthal direction and the Chebyshev collocation method in the radial direction ~\cite{RPeyret}.   The governing equations, given in full in Ref.~\cite{TDDM_simulation07},   are the 2D Navier-Stokes equation, the continuity equations for mass and charge, and one Maxwell equation for the nonlocal, coupled relationship between the charge and electric potential on the film.  We used the streamfunction and the vorticity as primitive variables. These were computed in the annular geometry $0 \leq \theta < 2\pi$ and $\alpha/(1-\alpha) \leq r \leq 1/(1-\alpha)$.
The boundary conditions are periodic in the $\hat{\theta}$ direction,  while a no-slip boundary condition holds at the edges of the annulus. 
% Numerical resolution:
To make sure the computations are well resolved, we employed up to $32$ Fourier modes, up to $40$ radial Chebyshev collocation points and a variety of time-step sizes.   We verified the simulation results using  different expansion orders and different time-step sizes~\cite{TDDM_simulation07}.

%We have previously validated the simulation code for the  zero shear case by calculating the dependence of ${\cal R}_c$ and the critical azimuthal mode number $m_c$ on $\alpha$ and  ${\cal P}$.  These were  benchmarked against previous theoretical results from linear stability analyses~\cite{DDM_PhyFluids_1999}, as discussed in ~\cite{TDDM_simulation07}.  Here,  we extend this study to the much richer case of sheared convection.  In contrast to the zero shear case, we find a more complex sequence of bifurcations which shows various oscillatory and localized vortex states and a clear route to chaos as the driving force is increased for fixed shear.
% ${\cal R}e$.
% toward the turbulent regime.  

% ----- suppression of the onset -----
Unlike in the case of 3D Taylor-Couette flow, here the 2D applied shear is itself stable and always stabilizes the conduction state, suppressing the onset of electroconvection.  At the primary bifurcation at the onset of convection,  
the critical azimuthal 
%
%  I added the word Fourier here, to satisfy the referee
%
%
Fourier mode with mode number $m_c$ becomes unstable.  
We denote ${\cal R}_c({\cal R}e=0) = {\cal R}_c^0$ and $m_c({\cal R}e=0) =  m_c^0$.
The simulation results for ${\cal R}_c^0$ and $m_c^0$ show excellent agreement with previous linear stability analyses~\cite{DDM_PhyFluids_1999,TDDM_simulation07}.  Under applied shear, we find
${\cal R}_c~>~{\cal R}_c^0$ and $m_c~\leq~m_c^0$.  This suppression of convection
is described by a reduced critical control parameter, $\tilde{\epsilon}=({\cal R}_c/ {\cal R}_c^0) -1$.  This reduced quantity can be extracted from experimental current data with essentially no adjustable parameters, and calculated using linear stability analysis~\cite{DDM_PhyFluids_1999}.  
The numerical simulation also produces values of the reduced Nusselt number $n = Nu - 1$ that can be directly compared to experimental data in the nonlinear regime.  We discuss such comparisons below.

% ----- primary supercritical bifurcation ------
In addition to the integrated quantities like $Nu$, the simulation 
%of course 
also provides directly the full vorticity, stream function, surface charge, and electric potential fields.  These basic fields are not directly accessible to experiment, except by using rather invasive flow visualization techniques which tend to unduly perturb the system~\cite{flow_vis}.  Thus, we can use the simulation to provide important new insights into the flow dynamics as ${\cal R}$ and ${\cal R}e$ vary, and our results shed new light on some previously unresolved experimental observations, while also making interesting new predictions.  

The simulations clearly show that the vicinity of the onset of convection under shear is much more complex than previously suspected, with new low-amplitude and oscillatory states predicted.
The simulations show that {\it all} primary bifurcations are continuous, or ``supercritical", even under shear. This is in contrast with previous experimental studies~\cite{
%zahir_prl, 
DDM_PRE_2001, ZDaya_PRE2002}  which 
%appeared 
were interpreted to show that the bifurcation became hysteretic and discontinuous, {\it i.e.}  ``subcritical", under a sufficiently large shear.   
This discrepancy may be traced to the fact that the excess current due to convection just above onset is very small, and can just barely be distinguished from the background conduction current due to noise and drift effects.  

\begin{figure}
\begin{center}
\includegraphics[width=3.1in]{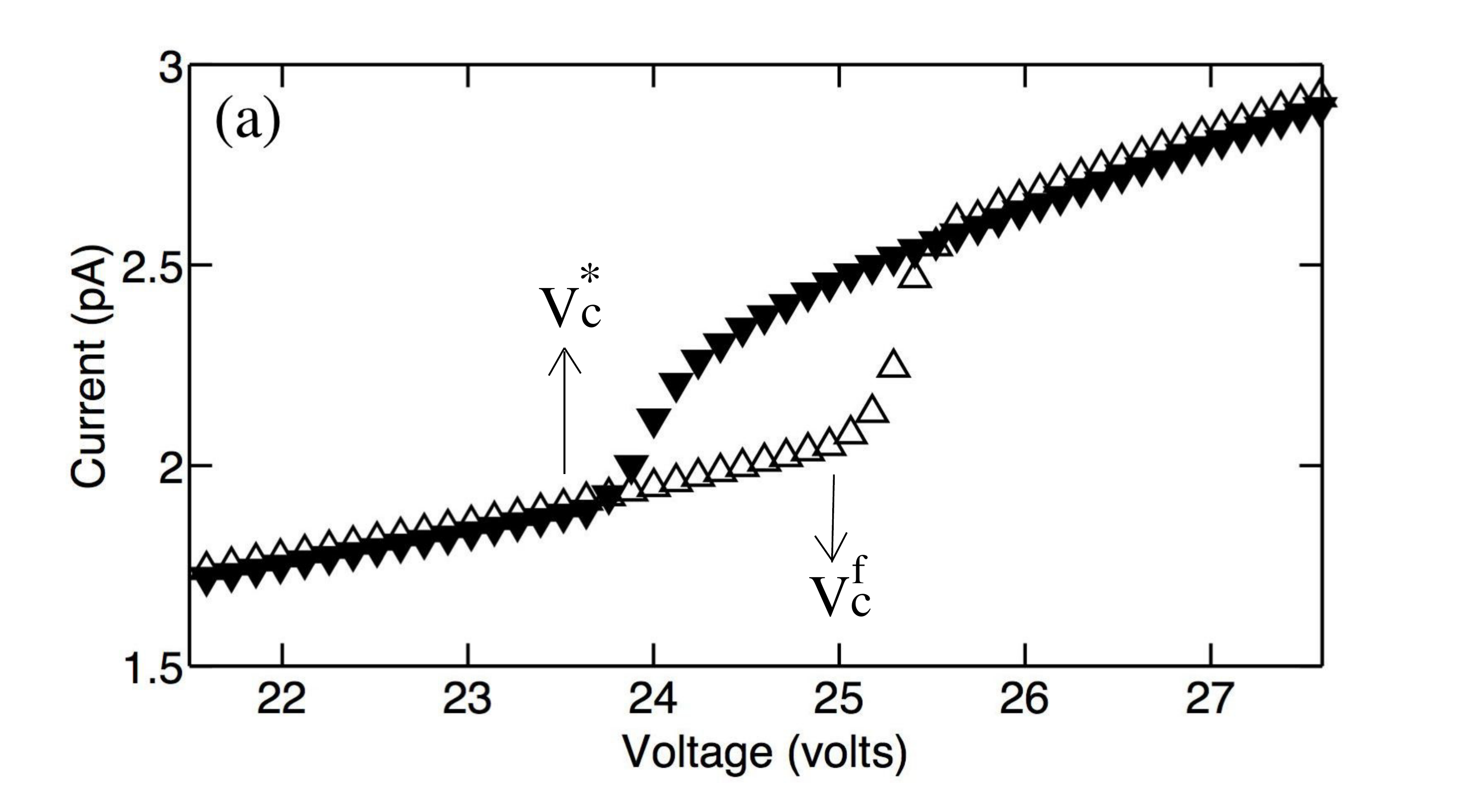}
\vspace{-2mm}
\includegraphics[width=3.1in]{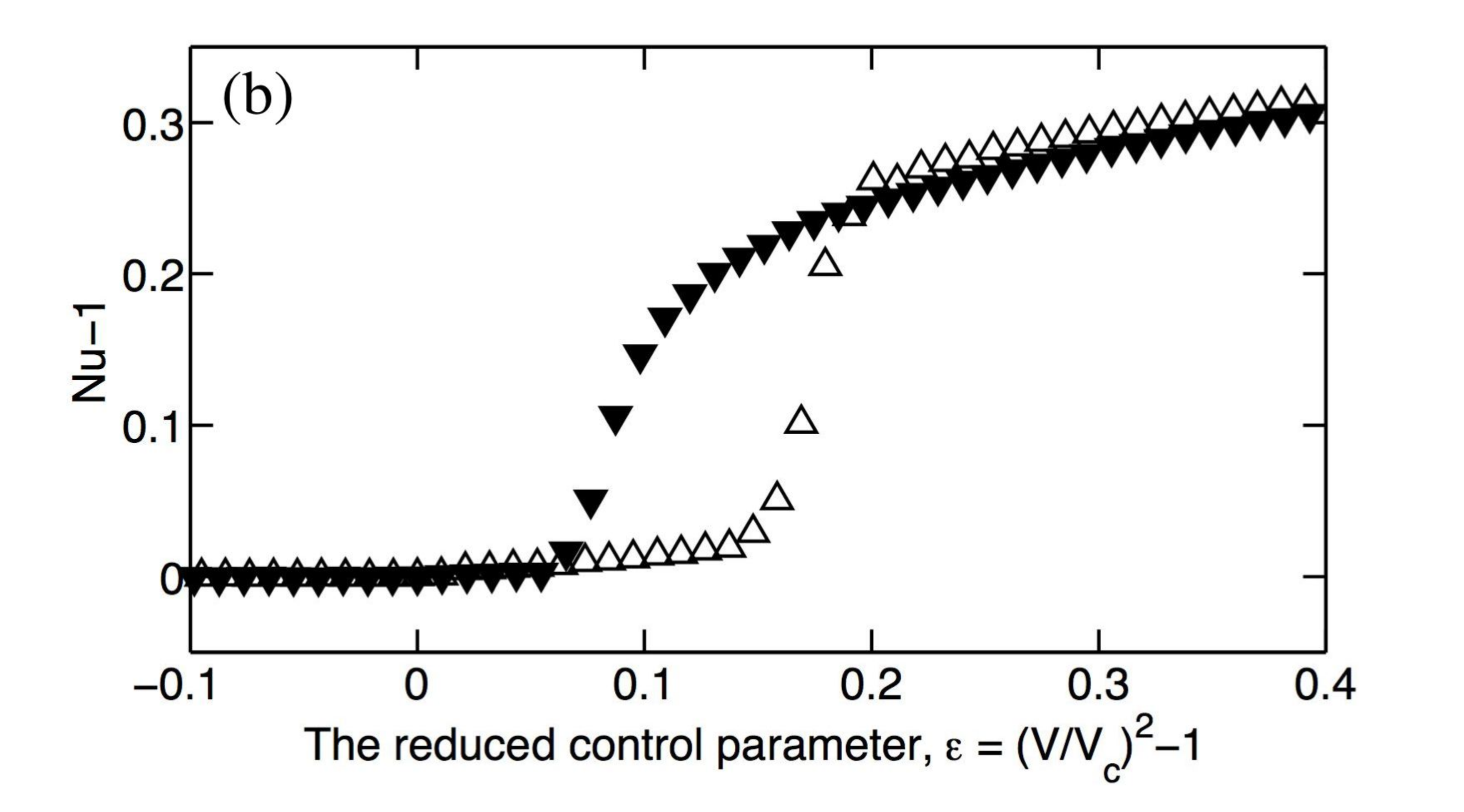}
\vspace{-2mm}
\includegraphics[width=3.1in]{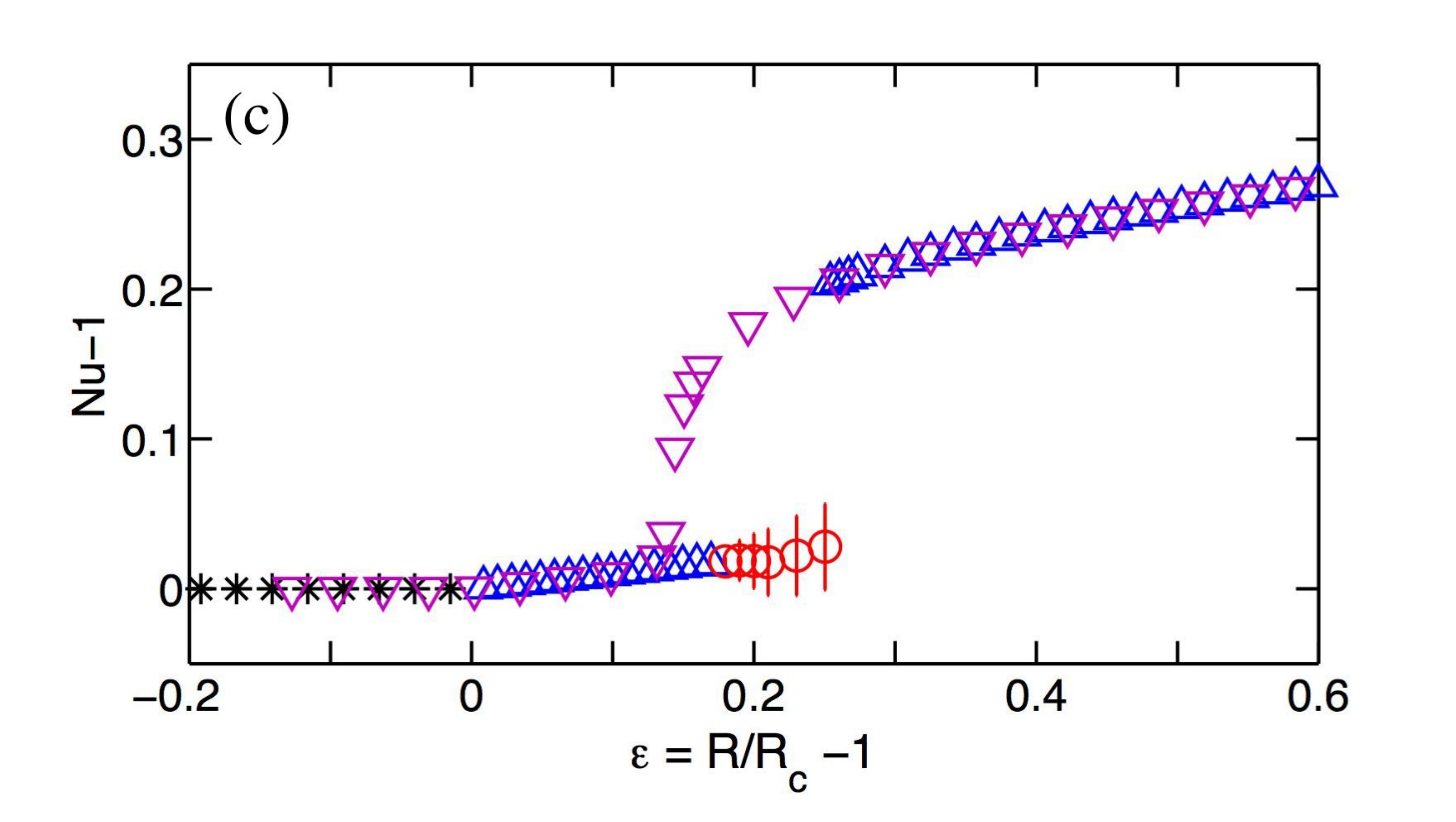}
\end{center}
\vspace{-5mm}
\caption{ (Color online) (a) Experimental raw current-voltage data from Ref.~\protect{\cite{ZDaya_PRE2002}} for $\alpha=0.56$, ${\cal P}=76,$ and ${\cal R}e=0.124$.  The open (filled) triangles are for increasing (decreasing) voltage. (b) The corresponding measured dimensionless convective current, $Nu-1$, {\it vs.} dimensionless control parameter $\epsilon = (V/V_c)^2-1$, extracted from the data in (a). (c) Simulation results for $Nu-1$ for the same parameters as in the previous experimental figure.  The film undergoes a sequence of bifurcations as ${\cal R}$ is slowly increased and then decreased ($\bigtriangledown$).  The sequence of flow regimes are conduction ($\ast$), steady rotating convection ($\bigtriangleup$), and oscillatory rotating convection ($\circ$), with the vertical bars showing the amplitude of oscillation.  A large hysteretic bifurcation is located at the jump in the dimensionless current that signals changes in the structure of the azimuthal Fourier modes.
%
% added Fourier here
%
}
\label{rr0_56_Re0_124}
\vspace{-4mm}
\end{figure}

Fig.~\ref{rr0_56_Re0_124}a shows raw experimental current-voltage data from Fig.~2 in Ref.~\cite{ZDaya_PRE2002} and the corresponding measured dimensionless quantities $Nu-1$ vs. ${\cal R}$. Fig.~\ref{rr0_56_Re0_124}c shows  the results of a full numerical simulation at the same dimensionless parameters as the previous figure.
 The integrated quantity $Nu - 1$ found numerically is in reasonable agreement with the experimental data.  However, an examination of the flow pattern 
 %and other fields 
 %
 %  from added 
 %
 reveals several new features which cannot be deduced from the experimental data alone.

Convection begins at a lower voltage than previously thought, and the first state encountered has a very low amplitude, corresponding to a small value of $Nu-1$.  In the analysis of experimental data like that shown in Fig.~\ref{rr0_56_Re0_124}a,  the critical voltage $V_c$ was misidentified to be near the jump at $V_c^f$ instead of somewhere close to $V_c^*$.  The numerical simulation shows that there is a sudden large increase in the current near $V_c^f$ which is associated with a sharp change in the mode structure, while the true onset of convection is lower, near $V_c^*$.   Just a trace of the low amplitude signal may be seen in the measured $Nu-1$ shown in Fig.~\ref{rr0_56_Re0_124}b.  In some unpublished data~\cite{zahir_thesis_p121}, the effect was
more pronounced than in Fig.~\ref{rr0_56_Re0_124}b, and was suspected to be due to low amplitude supercritical states, though the matter could not be experimentally resolved at that time.

The misidentification of $V_c$ means that the measured suppression of the onset on convection, represented by the experimental value of $\tilde{\epsilon}$, must be systematically too large.  The difference in $V_c^*$ and $V_c^f$ is typically of the order of the width of the hysteresis loop which is about $10 - 30\%$ of $V_c$, depending on other parameters.  
%
%  New sentence: reference added
%
The systematic overestimate of the experimental suppression may be seen in Fig.~7 of Ref.~\cite{DDM_PRE_2001}, where in the experimental data lie above the theoretical suppression from linear stability 
calculations.   

%- oscillatory convection \\
As ${\cal R}$ gradually increases beyond the steady sheared convection regime, the simulation shows that a Hopf bifurcation sets in leading to a periodic motion in a small ${\cal R}$ range, as shown in Fig.~\ref{rr0_56_Re0_124}c.  In addition to the usual rotation of the vortex pattern, 
%which leaves no signal in $Nu-1$, 
the amplitude becomes oscillatory in time, while generally still filling the circumference.  The steady sheared convection state does not show a time dependence in $Nu-1$ until these amplitude oscillations set in.

\begin{figure}
\begin{center}
\includegraphics[width=3.1in]{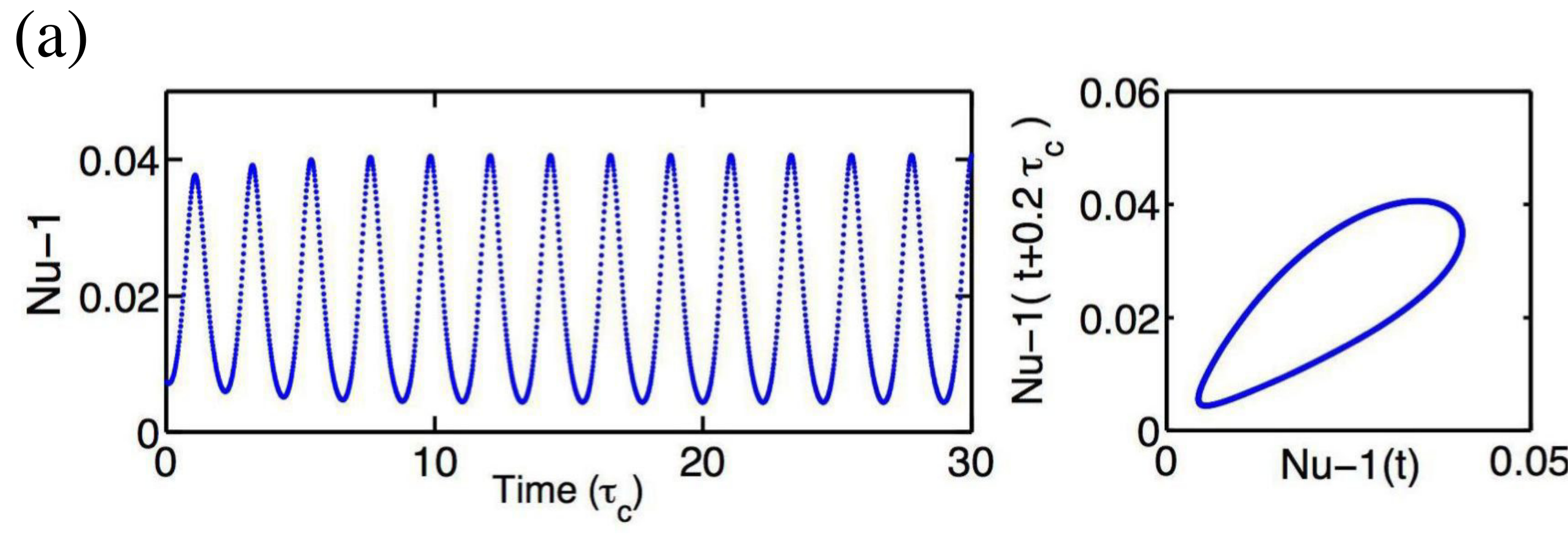}
\includegraphics[width=3.1in]{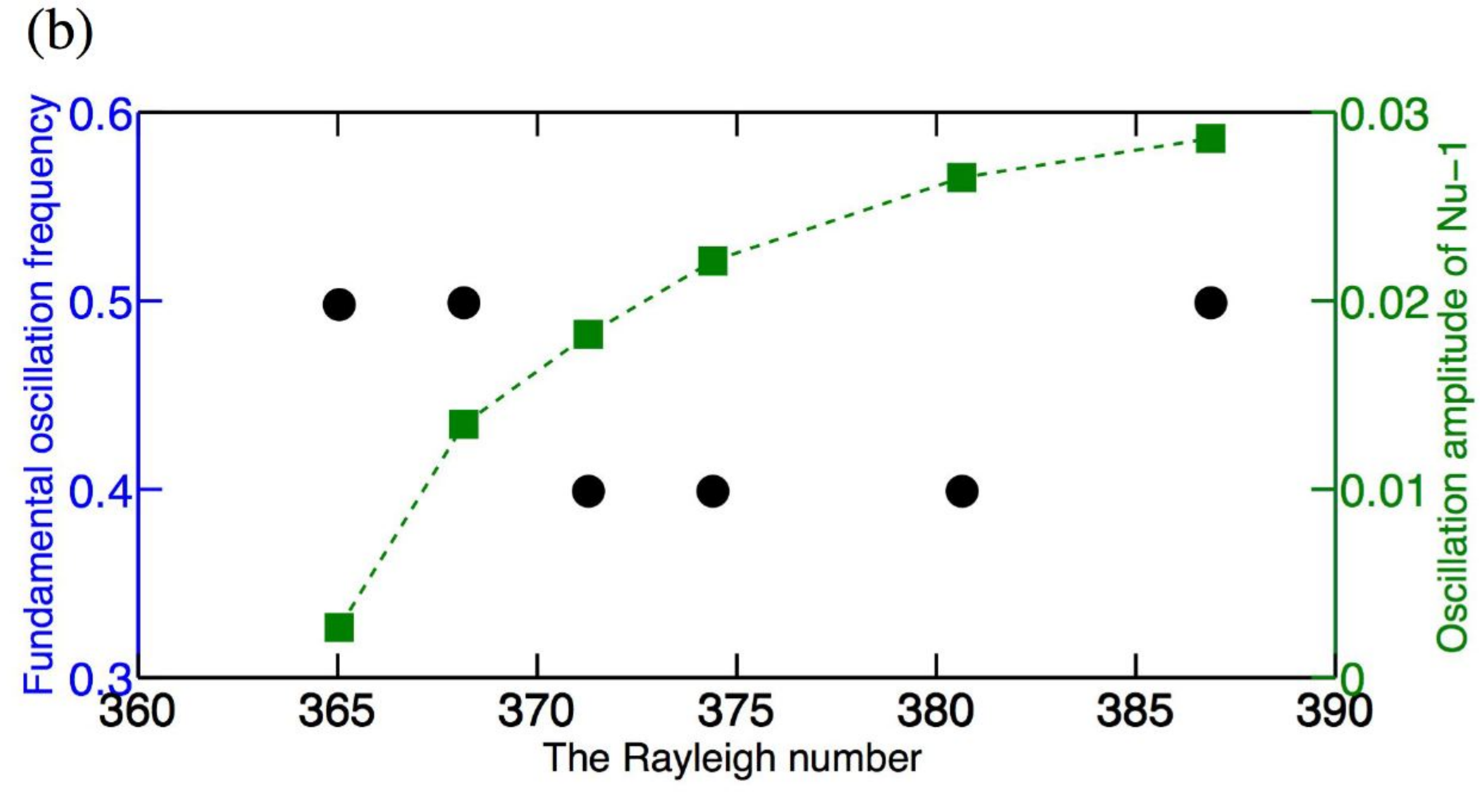}
\end{center}
\vspace{-3mm}
\caption{
(Color online) Numerical data for the oscillatory convective current $Nu-1$ for $\alpha=0.56$, ${\cal P}=75.8$, and ${\cal R}e=0.124$: (a) the time-series data for ${\cal R}=371.3$, together with the corresponding phase space trajectory, reconstructed using a time-delay method; (b) the fundamental frequencies ($\bullet$) and the oscillation amplitudes ($\blacksquare$) of $Nu-1$ as a function of ${\cal R}$.
}
\label{sheared_oscillatory_redNu_ts}
\vspace{-4mm}
\end{figure}

Oscillatory motion under a moderate shear corresponding to ${\cal R}e=0.124$ begins at about $1.18 \times {\cal R}_c$.  
Fig.~\ref{sheared_oscillatory_redNu_ts}a shows numerical time-series data of the oscillating convective current at ${\cal R}=371.3$.  After $\sim10\tau_q$, we find a one-frequency limit cycle is established.   The corresponding phase space reconstruction of this cycle is shown on the right of Fig.~\ref{sheared_oscillatory_redNu_ts}a, using the time-delay method with a sufficiently long time series of $Nu-1$.  The oscillatory regime spans the range $365< {\cal R} < 388$.  Fig.~\ref{sheared_oscillatory_redNu_ts}b shows the oscillation frequencies and amplitudes over this range. The oscillation frequency for ${\cal R}=371.3$ is $0.40$ in units of $\tau_q^{-1}$.   
Simultaneously, the rate of rotation of the vortex pattern itself, which is carried around by the imposed shear, measured in the range ${\cal R}=390$ --- $394$, was $2.36$~rad/$\tau_q$ counterclockwise, corresponding to a physical frequency of $0.38$~$\tau_q^{-1}$.  Thus, the frequencies of the amplitude oscillation and of the overall pattern rotation are comparable in the narrow window of oscillatory convection.     

%
%  added this detail back in here:
%

The oscillatory convection regime is not always present under shear, however.  We did not  observe oscillations for very small shears, such as ${\cal R}e=0.01$ and ${\cal R}e=0.06$, for similar supercritical values of ${\cal R}$.

Previous laboratory experiments~\cite{
%zahir_prl,
 DDM_PRE_2001, ZDaya_PRE2002} measured the time averaged total current passing through the film, producing data like that shown in Fig.~\ref{rr0_56_Re0_124}a.  
 %The time averaging was necessary to reduce noise in the very small currents being measured by an electrometer. 
 A re-examination of the time series that were averaged over, close to the sudden current increase that signals the end of the oscillatory regime, failed to unambiguously identify an oscillatory signal in the experimentally measured current.  The time series were not long enough to capture many periods of the oscillation, if it was present. The narrow window of voltages over which the oscillatory regime would exist was not sampled sufficiently.  New experiments, optimized for higher bandwidth and longer data acquisition of the current-voltage characteristics close to onset would be required to detect the oscillatory regime, supposing that it occurs as predicted by the simulation.

%- mode change dynamics \\
%- hysteresis Nu vs. Ra curve \\
The laboratory measurements unambiguously show a sequence of large subcritical bifurcations as a function of ${\cal R}$, for fixed ${\cal R}e$.  As ${\cal R}e$ is increased, these bifurcations evolve and proliferate in a regular way~\cite{ZDaya_PRE2002}.  Based on measurements of the current, and some crude flow visualization, it was supposed that these were jump transitions between homogeneous azimuthal Fourier 
%
%  added Fourier
%
modes of the form $m \rightarrow m \pm 1$.  The size of the changes in the current were consistent with this supposition.  Our numerical simulation of the same situation in which the flow patterns are revealed tells a different, more interesting story, however.

\begin{figure}
\begin{center}
\includegraphics[width=1.4in]{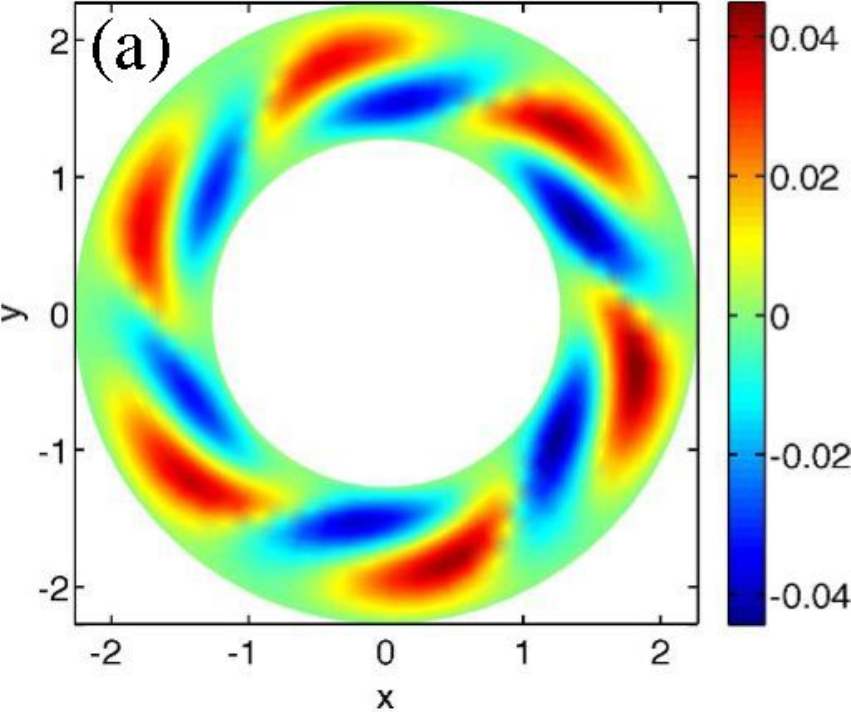}
\includegraphics[width=1.4in]{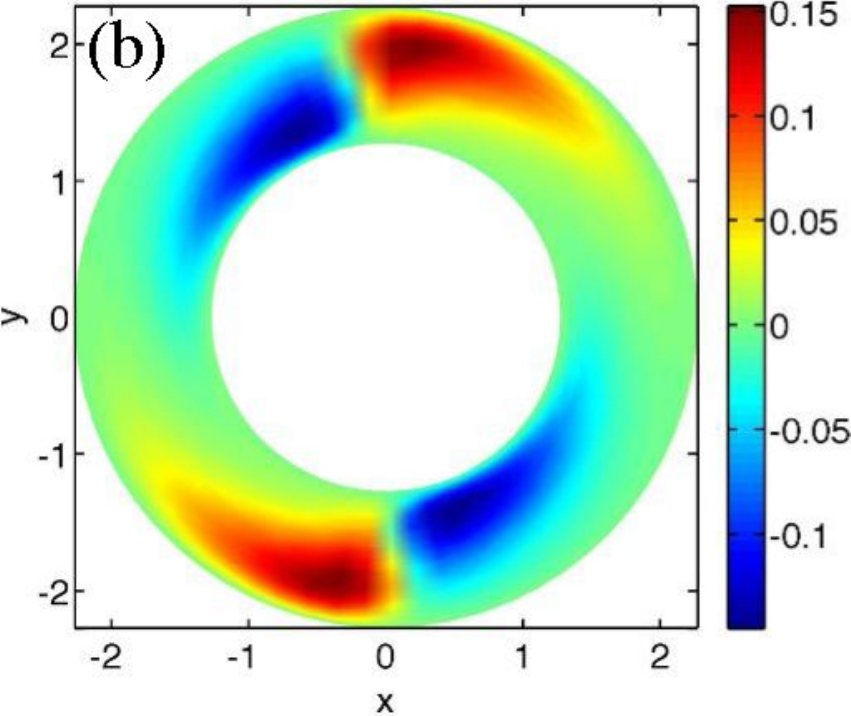}
\vspace{3mm}
\includegraphics[width=3.1in]{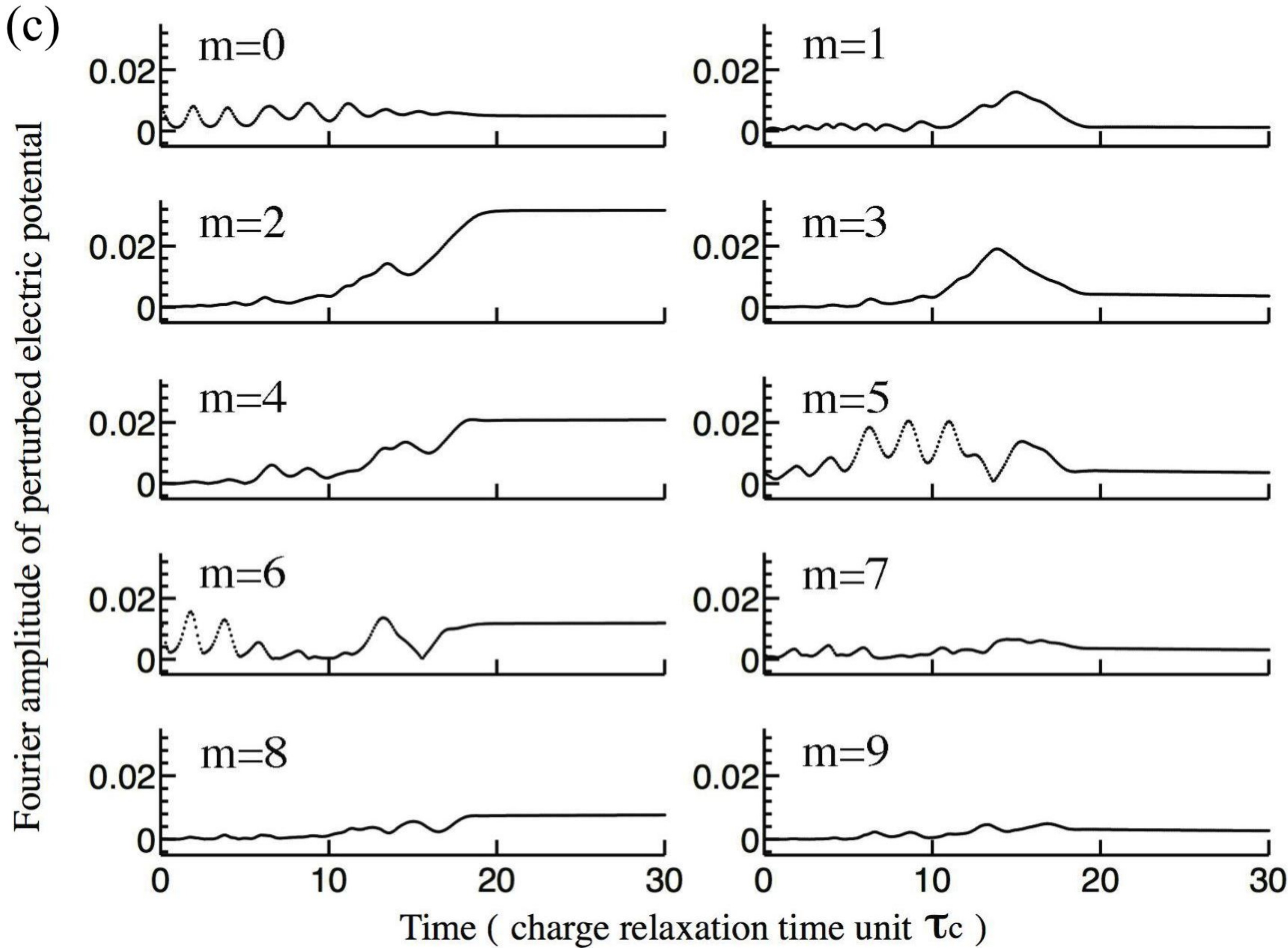}
\vspace{-3mm}
\caption{\label{mode_change} (Color online) The spatial patterns of the electric potential perturbation, ({\it i.e.} the total potential minus the conduction profile) at ${\cal R}=388.0$ and ${\cal R}e=0.124$, as in Fig.~\ref{rr0_56_Re0_124}c. (a) Initially, there are six oscillatory traveling vortex pairs.  (b)  After $30 \tau_q$, there are two non-oscillatory traveling vortices. (c) Time series of the Fourier amplitudes of the perturbed electric potential at midradius, showing the mode evolution during the transition.} 
\end{center}
\vspace{-8mm}
\end{figure}

\begin{figure}
\begin{center}
\includegraphics[width=2.9in]{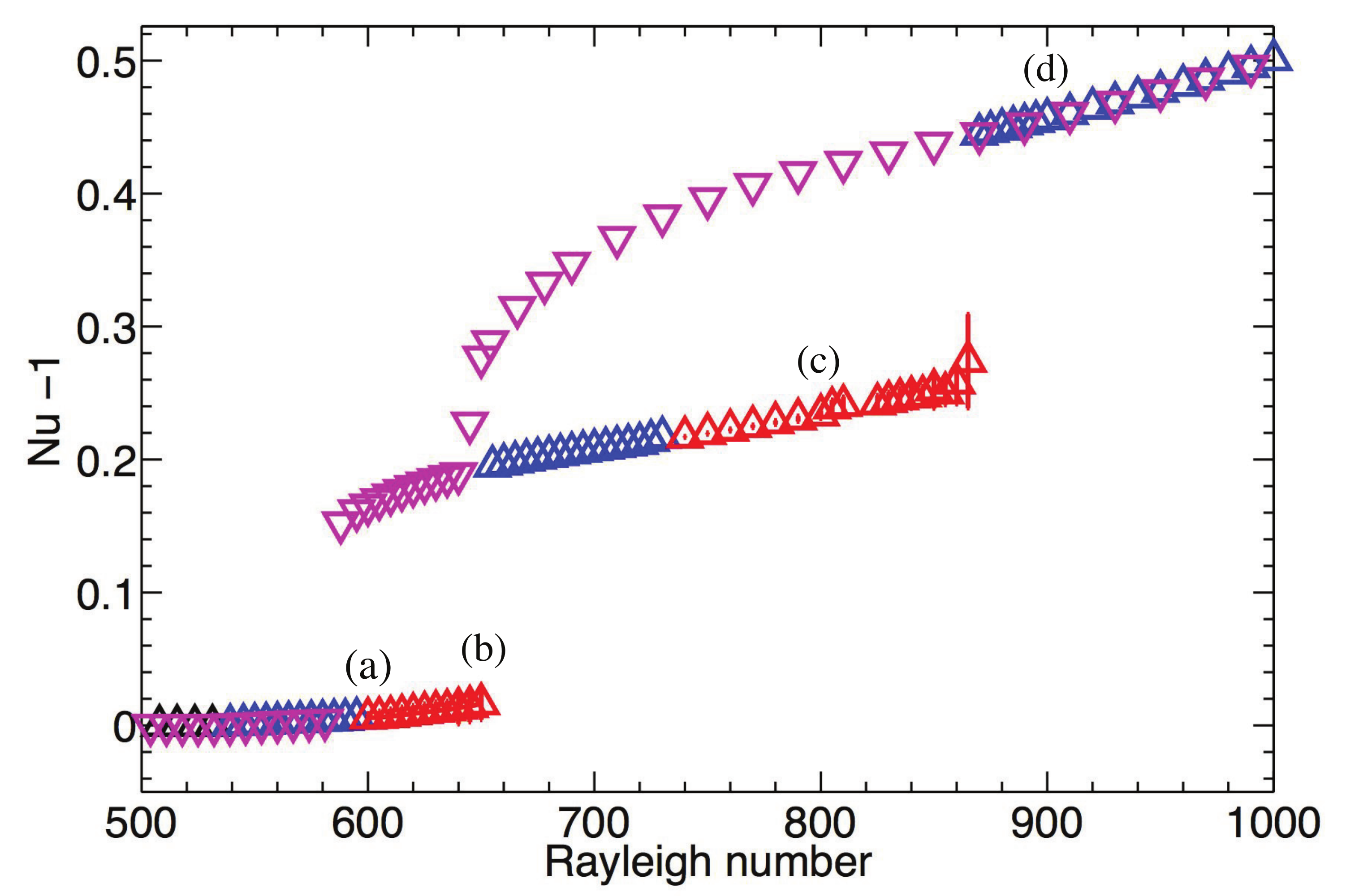}
\includegraphics[width=3.2in]{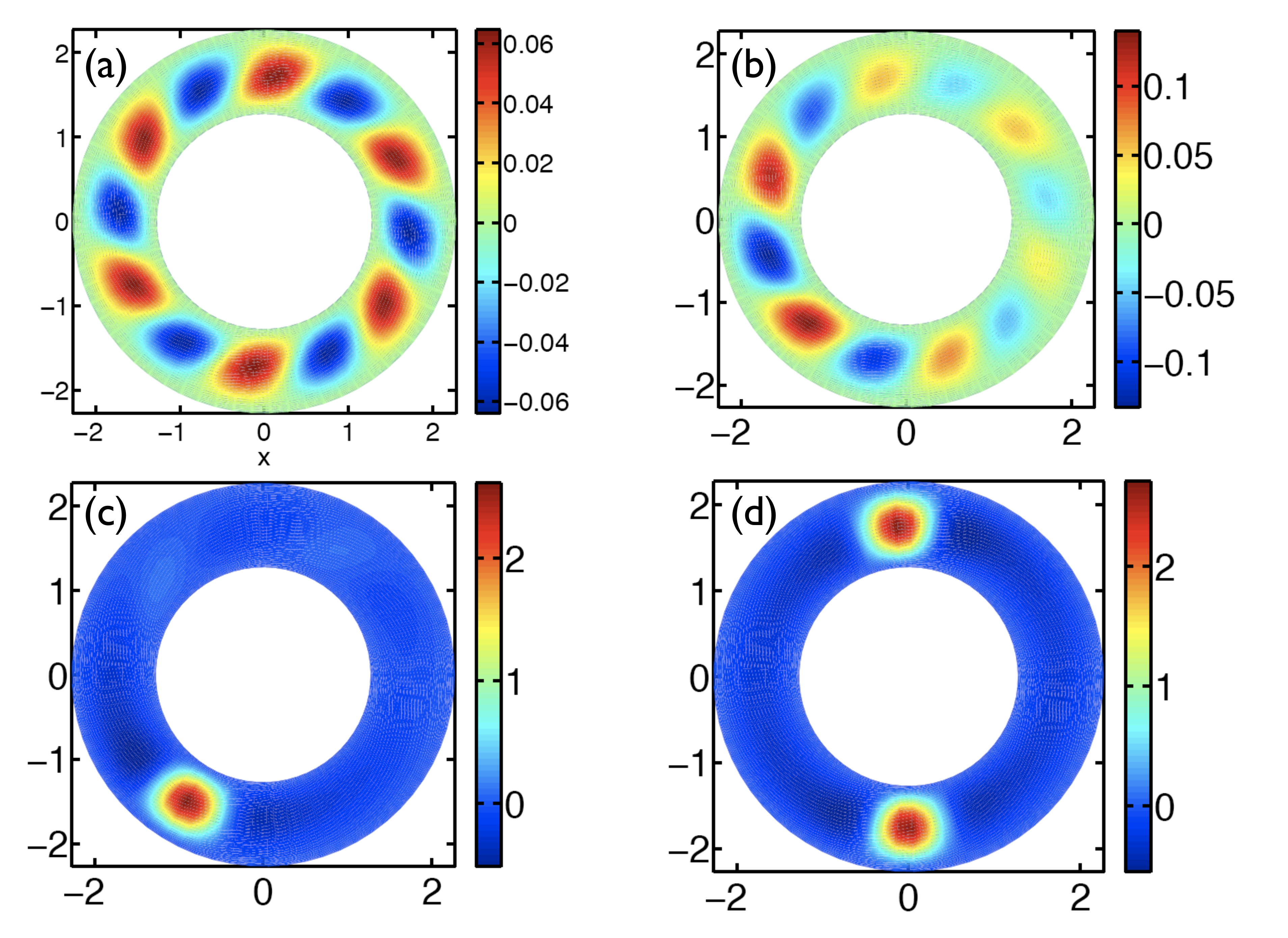}
\caption{\label{simulation_Re0_231}
(Color online) Numerical results for the dimensionless convective current, $Nu-1$, at $\alpha=0.56$, ${\cal P}=75.8$ and ${\cal R}e=0.231$.  In comparison with Fig.~\ref{rr0_56_Re0_124}c for which ${\cal R}e=0.124$, this larger shear produces more bifurcations over a similar range of the dimensionless control parameter.  The spatial patterns of the perturbed stream function shown below are marked by letters in the upper plot. (a) Oscillatory convection at ${\cal R}=600$ at $t=30~\tau_q$. (b) Undulating convection at a mode-changing bifurcation at ${\cal R}=650$ at $t=30~\tau_q$. (c) Oscillatory and localized convection, consisting of a single vortex at ${\cal R}=800$ at $t=20~\tau_q$. (d) Non-oscillatory, traveling localized convection at ${\cal R}=900$ at $t=20~\tau_q$.
}
\end{center}
\vspace{-8mm}
\end{figure}

As Fig.~\ref{mode_change}a and b show, the large subcritical transition shown in Fig.~\ref{rr0_56_Re0_124}c takes the traveling, oscillatory vortex pattern to a strongly localized state consisting of just two isolated vortices.  These vortices travel azimuthally in the direction of the shear, but are otherwise steady. 
%
%  refer forward to later fig to get the streamfunction
%
Fig.~\ref{mode_change}b shows the perturbed electric potential associated with the localized vortices. The fluid flow within the vortices resembles that shown in Fig.~\ref{simulation_Re0_231}d, which shows the perturbed stream function of a similar two vortex state under a larger shear.
 Fig.~\ref{mode_change}c shows the evolution of the mode amplitudes during the transition to the localized state.  $m=2$ and several harmonics combine to make up the localized vortices, indicating the broadband nature of the flow, which clearly requires strong mode coupling to remain coherent.
 
Upon reversing direction by decreasing ${\cal R}$ in the range of $342< {\cal R} < 388$, the convection exhibits hysteresis as shown in Fig.~\ref{rr0_56_Re0_124}c.  Time series data of the kinetic energy for decreasing ${\cal R}$ shows over-damped oscillations.  While decreasing ${\cal R}$ from $500$ down to $340$, the originally dominant Fourier modes at $m=2$ and $m=4$ decay to essentially zero and $m=6$ decays to a steady value.  On decreasing ${\cal R}$, the ${N}u-1$  curve shown in Fig.~\ref{rr0_56_Re0_124}c reaches what is presumably a saddle-node bifurcation endpoint, where it  rejoins the steady, non-oscillatory, $m=6$ convection regime just above the onset at ${\cal R}_c$.

Fig.~\ref{simulation_Re0_231} shows a simulation which captures two successive subcritical bifurcations under a larger shear.  In this case, the lower bifurcation results in a state consisting of a {\it single} traveling vortex.    The subsequent transition leads to a two vortex state, as before.  Several saddle nodes and new oscillatory localized states are seen.  These successive transitions in the simulation resemble the tree of bifurcations observed experimentally~\cite{ZDaya_PRE2002}.

Thus, we find numerically that the mode structure changes near the strongly subcritical  secondary bifurcations are much more complex than was apparent from the $IV$ data alone.  In particular, the modes may be oscillatory, and mode transitions can span more than $m \rightarrow m \pm 1$.  The traveling localized patterns above the bifurcation is not well characterized as having a single mode, and during the transition numerous coupled modes become active.  The localized states also carry orders of magnitude more current than the traveling and oscillatory states below the bifurcation --- so much so that the large subcritical jump was misidentified as the primary bifurcation in previous experimental studies~\cite{
%zahir_prl, 
DDM_PRE_2001, ZDaya_PRE2002}.

% new paragraph discussing localized states as snakes etc

Localized states in other systems~\cite{localized_theory_review} have sometimes been interpreted to be a consequence of heteroclinic cycles in the spatial coordinate~\cite{pomeau}  connecting a uniform background state to a spatially modulated state and back again.  Such states typically are stationary and occur in 1D systems with a reflection symmetry.  
% while our localized vortices travel.  
Here, such a symmetry is inherently strongly broken by the handedness of the applied shear.  In contrast, we find localized single vortices with only one sign of vorticity; they rotate internally in the same sense as the inner electrode, and travel azimuthally in the same direction.  The traveling aspect of the pattern may not be crucial, however, because it could be transformed away by passing to an appropriate rotating reference frame in which the localized state is stationary.  The dynamics in the 2D annulus is in fact unaffected by overall rotations (see Ref.~\cite{DDM_PhyFluids_1999} for a detailed explanation of this rather counterintuitive fact).

In the heteroclinic cycle model, the localized state can be thought of as two pinned fronts connecting states near the Maxwell point in a region of bistability between the two states.  The Maxwell point is the parameter value at which the ``energy'', or Lyapounov functional, of the two states is equal.  Near this point, the so-call {\it snaking} region contains localized solutions with various numbers of internal spatial modulations.  This picture has been applied to the case of localized ``convectons'' in binary fluid convection for example~\cite{snakes_bin_fluid_conv, convectons}, where the concentration and  temperature fields are respectively stabilizing and destabilizing, somewhat analogous to shear and applied electric potential in our case.  It is far from clear, however, whether this snaking picture can be relevant to the present case.  
We have not yet observed localized states consisting of multiple  counter-rotating vortices, only isolated vortices of a single handedness surrounded by regions of very weak reverse vorticity.
Also, the range of existence of the localized states appears to span a range of parameters too wide to be confined to a narrow snaking region.  They emerge not from a uniform state, but rather from oscillatory low-amplitude precursor patterns. Of course, neither the experiment, nor the time-stepping simulation, give us full access to all the solutions, stable and unstable, that may exist in the vicinity of the bifurcation.    It is also worth noting that while the simulation reproduces the experimentally measured currents quite well, the localized states have not been observed directly using the limited flow visualization available.

\begin{figure}
\begin{center}
\includegraphics[width=1.7in]{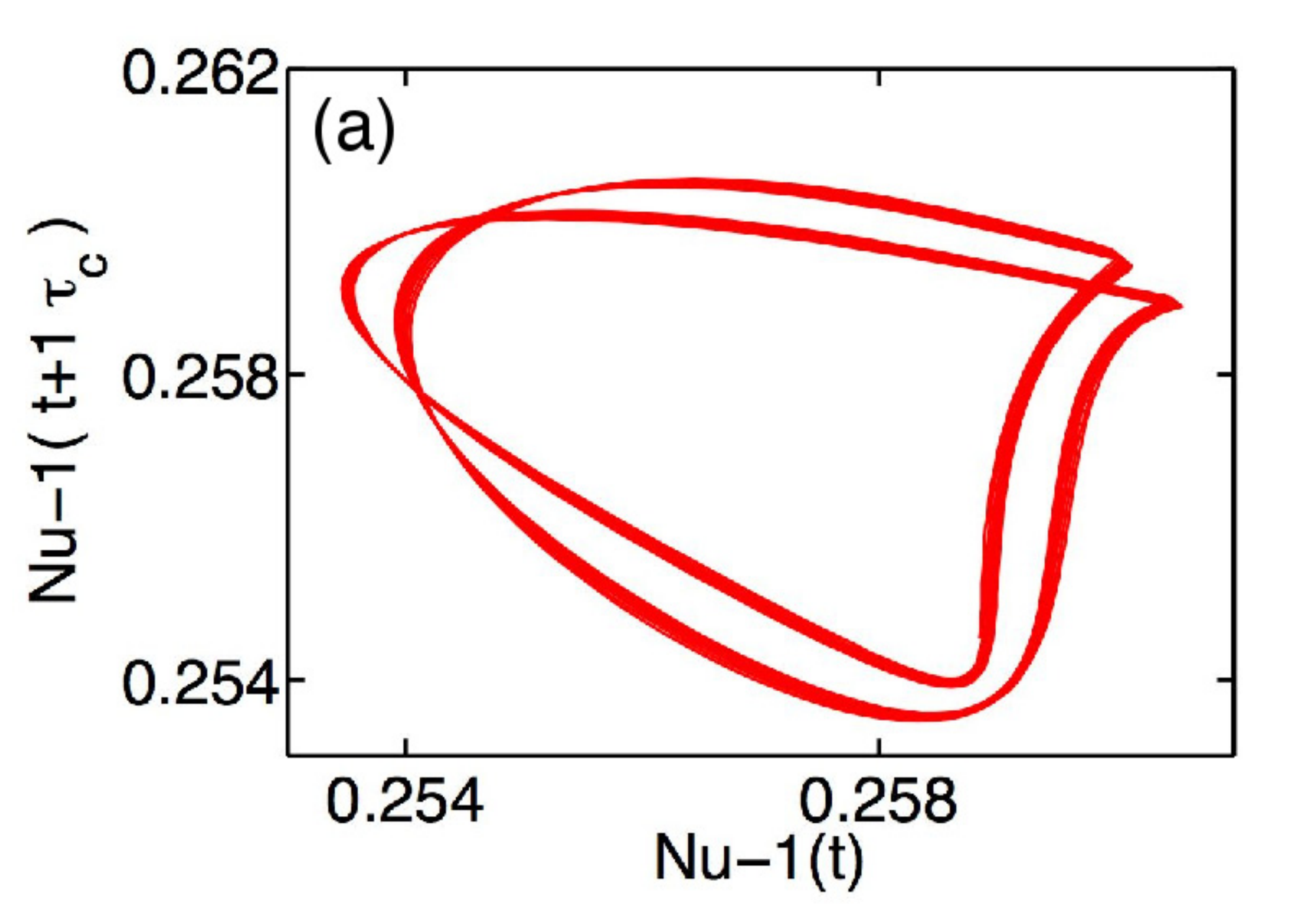}
\includegraphics[width=1.7in]{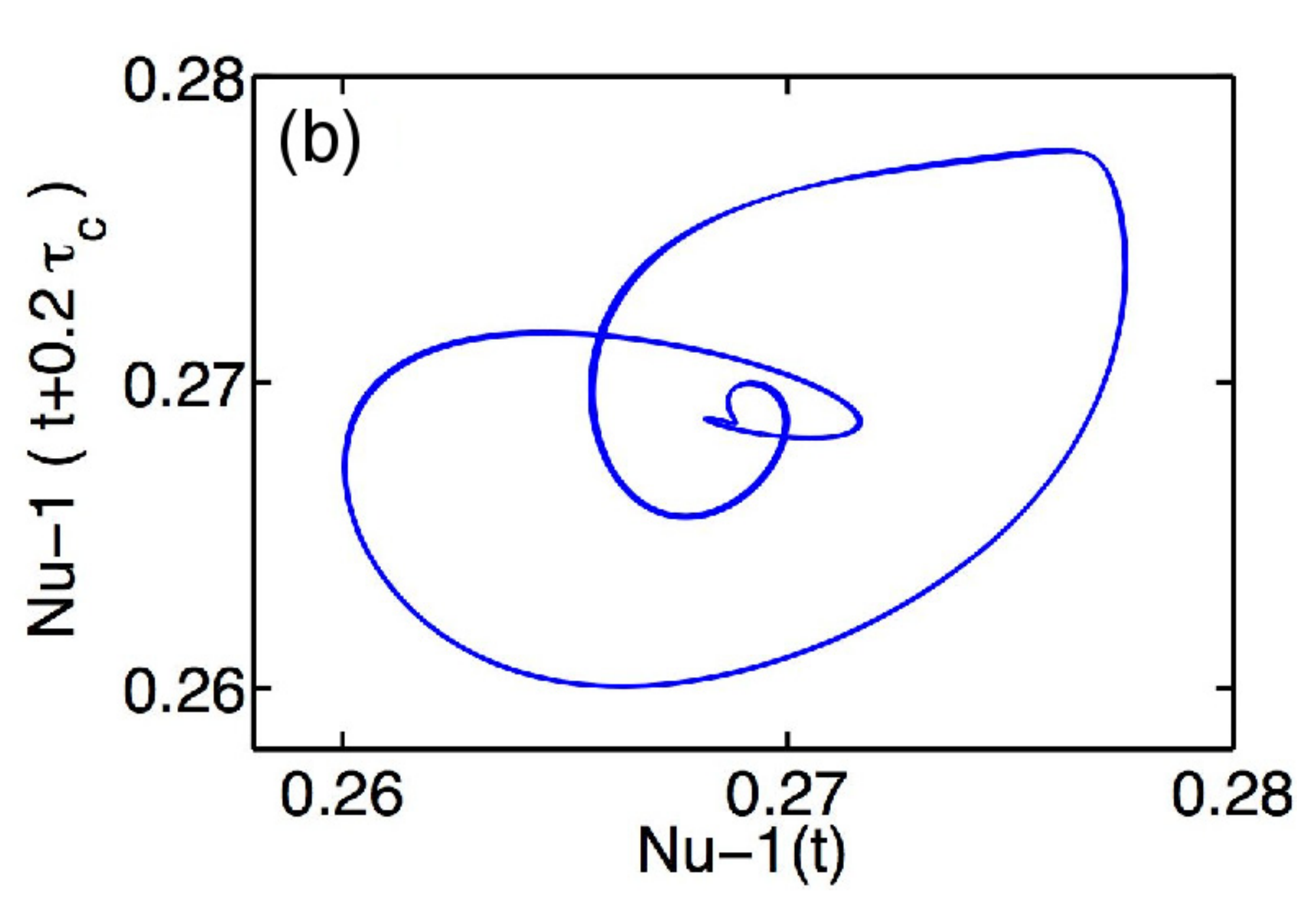}
\end{center}
\caption{\label{route_to_chaos_with_shear}
(Color online) Projections of the system attractor of $Nu(t)-1$ using a time-delayed reconstruction method in sheared convection with $\alpha=0.47$,  ${\cal P}=16.3$ and a large ${\cal R}e=0.8$. (a) 2-frequency quasiperiodic flow (2-torus) at ${\cal R}=639.3$. (b)  3-frequency quasiperiodic flow (3-torus) at ${\cal R}=673.4$. These are spatially localized states which exhibit multifrequency oscillations in the time domain as the driving force is increased toward to a chaotic state.}
\vspace{-4.5mm}
\end{figure}

% - route to chaos in sheared convection\\
For larger ${\cal R}e$ and ${\cal R}$, we observe the onset of chaotic flow, which followed the Ruelle-Takens-Newhouse scenario for the route to chaos~\cite{RTN_route_to_chaos}.   Fig.~\ref{route_to_chaos_with_shear} shows the time domain of $Nu-1$ for $\alpha=0.47$,  ${\cal P}=16.3$, and the large shear ${\cal R}e=0.8$.  With increasing ${\cal R}$,  we first find steady convection,  then a periodic state (a period-1 limit cycle), as in Fig.~\ref{sheared_oscillatory_redNu_ts}, followed by a two-frequency quasiperiodic flow (a 2-torus), a three-frequency quasi-periodic flow (a 3-torus), and finally chaos.  Fig.~\ref{route_to_chaos_with_shear} illustrates the attractors of those states using a time-delay reconstruction~\cite{book_on_chaotic_data} from time-series of $Nu-1$.  
%The suppression of the convection onset in this strongly sheared simulation is $\tilde{\epsilon}=3.96$, compared to the theoretical prediction of $\tilde{\epsilon}=3.6$, 
%
%The state of steady convection exhibits five traveling vortex pairs which agrees with theoretical prediction of $m_c=5$. 
%
 We first find a 1-frequency periodic motion at ${\cal R}=485.6$ with a basic frequency of $f_1\sim 0.60~\tau_q^{-1}$. Subsequently, the basic frequency of the periodic state changes slightly in the range $0.40$ --- $0.50~\tau_q^{-1}$ as ${\cal R}$ increases. The 2-torus motion at ${\cal R}=639.3$ has fundamental frequencies $f_1=0.70~\tau_q^{-1}$ and $f_2=0.80~\tau_q^{-1}$. In space,  the flow is localized to a single traveling vortex like that shown in Fig.~\ref{simulation_Re0_231}c.
 The 3-torus state at ${\cal R}=673.4$ is also localized and has the basic frequencies $f=0.31, 0.69$, and $1.31$,  in units of  $\tau_q^{-1}$.  Finally,  at higher ${\cal R}=1190$, chaotic motion is observed, yet the traveling state remains localized in space. The complex time dependence comes from small blobs of charge which are chaotically emitted from the electrodes near the vortex.

% CONCLUSION PARAGRAPH\\
In conclusion, direct numerical simulation of sheared electroconvection provides an interesting new window on localized states, bifurcations and chaos in 2D fluid flows.  It nicely complements previous experimental studies, which were mainly limited to current measurements, by allowing visualization of the basic fields. The simulations revealed an unexpected low amplitude state just above the onset of convection, and oscillatory states which have not yet been  systematically investigated in experiment.  A strongly subcritical bifurcation makes transitions between these oscillatory states to novel new states with localized  traveling vortices.  These carry much more current.  Their localization is presumably due to some sort of mode interaction, but the exact nature of their stability is unknown.  New, more sensitive experiments with better visualization will be necessary in order to study these states experimentally.  Under stronger electrical forcing and higher rates of shear, we also observed a transition to localized chaotic convection {\it via} a Ruelle-Takens-Newhouse scenario.  The relatively simple geometry of this system makes it an ideal place to explore ideas about localized, chaotic and turbulent states both theoretically and experimentally.

%More generally, this 2D system with naturally periodic boundary conditions, and for which forcing and shear are independently controllable, will be an interesting place to examine recent ideas on higher dimensional invariant manifolds on the way to turbulence~\cite{cvit, 
%%RBraun_PRE98, FFeudel_PRE95, 
%Heijst_PRL05}
%%, quasi_2D_exp_JFM1986}.

\acknowledgments
We gratefully thank V. B. Deyirmenjian for the stimulating and helpful discussion, and J. H. P. Dawes and J. Burke for comments. We also  thank the Canadian Institute of Theoretical Astrophysics (CITA) for the use of computational facilities. This research was supported by the the Natural Sciences and Engineering Research Council (NSERC) of Canada.

\end{document}